\documentclass[prl,nofootinbib,superscriptaddress]{revtex4}
\usepackage[a4paper,left=1.5cm,right=1.5cm,top=3cm,bottom=3cm]{geometry}
\usepackage[toc]{appendix} 

\pdfoutput=1
\usepackage{amssymb,amsmath,amsfonts}
\usepackage[dvipsnames]{xcolor}
\usepackage{graphicx}
\usepackage{longtable}
\usepackage{verbatim}
\usepackage{color}
\usepackage{mdframed}
\usepackage{soul}
\usepackage{mathrsfs}
\usepackage{amsfonts,amssymb,mathrsfs,amsmath,esint}
\usepackage{slashed, cancel}
\usepackage{framed}
\usepackage{mdframed}
\usepackage{simplewick} 
\allowdisplaybreaks

\usepackage{latexsym}
\usepackage{graphicx}
\graphicspath{{./Figures/}}
\usepackage[dvipsnames]{xcolor}
\usepackage{booktabs}
\usepackage{datetime}
\newdateformat{mydate}{\THEDAY{ }\monthname[\THEMONTH]{ }\THEYEAR}

\usepackage{tikz}
\usepackage{color}
\usepackage{framed}
\usepackage{hyperref}
\hypersetup{colorlinks, citecolor=bluscuro, linkcolor=black, urlcolor=bluscuro}
\definecolor{rossos}{cmyk}{0,1,1,0.55}
\definecolor{bluscuro}{rgb}{0.15, 0.2, .85}
\definecolor{bluchiaro}{cmyk}{1,.3,0.,0.1}

\graphicspath{{./Figures/}}

\newcommand{\be}{\begin{equation}}
\newcommand{\ee}{\end{equation}}
\newcommand{\bea}{\begin{eqnarray}}
\newcommand{\eea}{\end{eqnarray}}
\newcommand{\beq}{\begin{equation}}
\newcommand{\eeq}{\end{equation}}
\newcommand{\lp}{\left(}
\newcommand{\rp}{\right)}
\newcommand{\llp}{\left[}
\newcommand{\rrp}{\right]}
\def\beqa{\begin{eqnarray}}

\def\pk{{\text{\tiny pk}}}
\def\TT{{\text{\tiny TT}}}
\def\H{{\text{\tiny H}}}

\def\ta{{\text{\tiny ta}}}
\def\GW{{\text{\tiny GW}}}

\def\vx{{\vec{x}}}

\def\d{{\rm d}}

\def\pk{\text{\tiny pk}}

\newcommand{\Pz}{\mathcal P_\zeta}

\newcommand{\dd}{{\rm d}}
\def\eeqa{\end{eqnarray}}

\def\lsim{\mathrel{\rlap{\lower4pt\hbox{\hskip0.5pt$\sim$}}
    \raise1pt\hbox{$<$}}}         
\def\gsim{\mathrel{\rlap{\lower4pt\hbox{\hskip0.5pt$\sim$}}
    \raise1pt\hbox{$>$}}}         

\def\lsim{~\rlap{$<$}{\lower 1.0ex\hbox{$\sim$}}}
\def\bsim{~\rlap{$>$}{\lower 1.0ex\hbox{$\sim$}}}

\def\la{\langle}
\def\ra{\rangle}
\def\dd{{\rm d}}
\def\ln{{\rm ln}}

\usepackage[normalem]{ulem}

\newcommand{\arXiv}[2]{\href{http://arxiv.org/pdf/#1}{{\tt [#2/#1]}}}
\newcommand{\arXivold}[1]{\href{http://arxiv.org/pdf/#1}{{\tt [#1]}}}

\numberwithin{equation}{section}
\renewcommand\theequation{\arabic{section}.\arabic{equation}}

\begin{document}

\title{Gravitational Waves from  Peaks}
\author{V. De Luca}
\affiliation{D\'epartement de Physique Th\'eorique and Centre for Astroparticle Physics (CAP), Universit\'e de Gen\`eve, 24 quai E. Ansermet, CH-1211 Geneva, Switzerland}
\author{V. Desjacques}
\affiliation{Physics department and Asher Space Science Institute, Technion, Haifa 3200003, Israel}
\author{G. Franciolini}
\affiliation{D\'epartement de Physique Th\'eorique and Centre for Astroparticle Physics (CAP), Universit\'e de Gen\`eve, 24 quai E. Ansermet, CH-1211 Geneva, Switzerland}
\author{A. Riotto}
\affiliation{D\'epartement de Physique Th\'eorique and Centre for Astroparticle Physics (CAP), Universit\'e de Gen\`eve, 24 quai E. Ansermet, CH-1211 Geneva, Switzerland}
\affiliation{CERN, Theoretical Physics Department, Geneva, Switzerland}

\date{\today}

\begin{abstract}
\noindent
We discuss a novel mechanism to generate gravitational waves in the early universe. A standard way  to  produce primordial black holes   is to enhance  at small-scales the overdensity  perturbations generated during inflation. The  latter, upon horizon re-entry,  collapse into black holes. They must be  sizeable enough and are therefore associated to rare peaks. There are however  less sizeable and much less rare  overdensity peaks which do not   end up forming  primordial black holes and  have a non-spherical shape.  Upon collapse, they  possess  a time-dependent non-vanishing mass quadrupole which gives rise to the generation of gravitational waves.  By their nature, such gravitational waves are complementary to   those  sourced at second-order by the very same  scalar perturbations responsible for the formation of the primordial black holes. 
Their amplitude is nevertheless typically about two orders of magnitude smaller and therefore hardly measurable.
\end{abstract}

\maketitle


\section{I. Introduction}
\setcounter{section}{1}

\noindent
A new era, dubbed gravitational wave astronomy, has begun since the  detection of gravitational waves (GWs) produced by the  merging of two $\sim 30\, M_\odot$ black holes \cite{ligo1}. At the same time, the very same discovery has given fresh life to the physics of primordial black holes (PBHs) and, more in particular, to the prospect that all (or a significant part of) the dark matter of the universe is in the form of primordial black holes \cite{bird,juan} (see Ref. \cite{revPBH} for a review and more references therein). 

A standard scenario to produce primordial black holes in the early universe is to generate during inflation  a large power spectrum of the scalar comoving curvature perturbation,  ${\cal P}_\zeta\sim 10^{-2}$ on small scales  \cite{s1,s2,s3}, i.e. greater (by at least seven orders of magnitude) than  the large-scale power spectrum which is ultimately responsible for the generation of the CMB anisotropies as well as the seeds of the large-scale structure.   

Such small-scale  sizeable  perturbations are subsequently   transferred  to the radiation fluid  via the reheating process, which transforms the vacuum energy responsible for inflation into relativistic degrees of freedom. This gives rise  to  primordial black holes if the curvature perturbations are  large enough upon horizon re-entry. In particular, a 
  region  collapses to  a primordial black hole  if  the overdensity  is larger than a critical value $\delta_{\rm c}\lesssim 1$. 
  
 In this picture primordial black holes may be thought as originating  from peaks of the density contrast $\delta(\vec x)=(4/9a^2H^2)\nabla^2\zeta(\vec x)$, that is from maxima of the local overdensity, where $H$ is the Hubble rate and $a$ the scale factor.  The average density of maxima of a general three-dimensional Gaussian field can be calculated as a function of heights of the maxima, which we call $\delta_\pk$, through peak theory \cite{bbks}~\footnote{The overdensity $\delta$ is not really a Gaussian field as it is related to the Gaussian, comoving curvature perturbation by a nonlinear relation. Therefore, the overdensity statistics is unavoidably non-Gaussian \cite{ng1,ng2,ng3}. Nevertheless, this point will not alter our conclusions.}.
 
In order for the primordial black holes to form a non-negligible fraction of the dark matter, one requires the mass fraction to be \cite{bbks}
 \be
 \label{beta}
\beta \simeq 6 \cdot 10^{-9} \lp \frac{M}{M_\odot} \rp^{1/2} \simeq 10^{-2} \nu^2 e^{-\nu^2/2},
 \ee
 where we have introduced the rescaled peak's height $\nu = \delta_{\pk}/\sigma_0$ as a function of the variance of the overdensity
\be
\sigma^2_0 =\int \frac{{\rm d}k}{k}
\,W^2(k,R_H)
\,{\cal P}_{\delta}(k).
\ee
Here  ${\cal P}_\delta$ is the overdensity power spectrum, $R_H$ is  the comoving horizon length  $R_H=1/aH$  and $W(k,R_H)$ is a  window function. From Eq. \eqref{beta} one deduces that the peak's height must be large compared to $\sigma_0$, 
 \be
 \nu=\frac{\delta_{\pk}}{\sigma_0}\gsim {\cal O}(6\div 8 ).
 \ee  
This means that only very rare peaks end up forming primordial black holes. 

 What about those peaks of the small-scale overdensity power spectrum, with an amplitude not large enough to end up as primordial black holes~? Do they imprint any observational signature~?
 
 In this paper we argue that they do under the form of gravitational waves. The reason is simple. It is well-known that a self-gravitating system emits GWs if it possesses a non-vanishing and time-dependent quadrupole \cite{maggiore}. Peaks with moderate values of the amplitude are not spherical but, rather, ellipsoidal objects, that is they possess a quadrupole. Furthermore, when the comoving Hubble radius grows and becomes of the order of the size of the peak, these overdense perturbations decouple from the background
 and contract, leading to the generation of GWs.

 To the best of our knowledge, this is a  novel mechanism, which should not be confused with the generation of GWs at second-order in perturbation theory by the  comoving curvature  perturbations at horizon re-entry  \cite{saito,e,gbp,noi1,noi2}. For instance, a distinctive feature of the mechanism we describe in this paper is that the source of GWs disappears in the limit of spherical peaks. In the rest of the paper, we describe the mechanism in more details.
 
The paper is organised as follows. In Section II we start from peak theory to describe the origin of the mass quadrupole. In Section III we compute the amount of GWs sourced by the quadrupole and in Section IV we discuss the results, making a comparison with the  second-order contribution. Section V summarises our conclusions. Finally, a couple of Appendices provide additional details about the calculation. 
\bigskip

\noindent
\section{II. The origin of the mass quadrupole}
\setcounter{section}{2}
\setcounter{equation}{0}
\noindent
Our starting point is peak theory \cite{bbks}. Consider a generic peak of the small-scale density fluctuations. We expand the overdensity around the peak up to second-order (in spatial derivatives)
\begin{align}
\delta(\vx) = \frac{ \rho(\vx) -\bar\rho}{\bar\rho} \simeq \delta_\pk + \frac{1}{2}\zeta_{ij}(x- x_\pk)^i (x- x_\pk)^j,
\qquad
 \zeta_{ij} = \left.\frac{\partial^2 \delta}{ \partial x^i \partial x^j}\right|_\pk,
\end{align}
where $\vx$ are comoving coordinates, the gradient term vanishes since we are dealing with local density maxima, $\zeta_{ij}$ is evaluated at $x_\pk$ and $\bar\rho$ is the average energy density.
We will ignore higher-order terms and, therefore, approximate the isodensity contours of the perturbation by concentric ellipsoids.
Rotating the coordinate axes such that they are aligned with the principal axes of the constant-overdensity ellipsoid,
we can write the expansion of the overdensity around the peak in terms of the eigenvalues $\lambda_i$ of the matrix $-\zeta_{ij}/\sigma_2$ as
\be
\label{alambda}
\delta(\vx) \simeq \delta_\pk -\frac{1}{2}\sigma_{2} \sum_{i=1}^{3}\lambda_i (x^i- x_\pk^i)^2.
\ee
Here, $\sigma_{2}$ is the characteristic square root variance of the components of $\zeta_{ij}$, and we use from now on the standard definition
\be
\sigma^2_j =\int \frac{{\rm d}k}{k}\,k^{2j}
\,W^2(k,R_H)
\,{\cal P}_{\delta}(k).
\ee
Eq. (\ref{alambda}) can be rewritten as 
\be
\frac{2(\delta - \delta_\pk)}{\sigma_{2}} = - \sum_{i=1}^{3}\lambda_i (x^i- x_\pk^i)^2.
\ee
We take the boundary of the peak to be the ellipsoid of overdensity $\delta = \alpha \delta_\pk$, with $\alpha<1$. The previous relation yields
\be
2\frac{\sigma_0}{\sigma_{2}}(1-\alpha)\nu =  \sum_{i=1}^{3}\lambda_i (x^i- x_\pk^i)^2,
\ee
the solution of which defines the boundary of the peak. The principal semi-axes of this ellipsoid are given by
\be
\label{semiaxes}
a_i^{2} = 2\frac{\sigma_0}{\sigma_2}\frac{(1-\alpha)}{\lambda_i}\nu.
\ee
While for peaks ending up in PBHs $\alpha$ is estimated to be $1/5$ \cite{spin}, here and henceforth we take care of peaks which have a positive overdensity, and thus we will make the approximation $1-\alpha\simeq 1$.
We turn now to the mass quadrupole defined as
\begin{equation}
	Q^{ij} = M^{ij} - \frac{1}{3} \delta^{ij} M^k_k,
\end{equation}
where ($r^i=a x^i$ are the physical coordinates) 
\begin{align}
	M^{ij} &= \int_{V_{\rm e}}\!\!\d^3r\, r^i r^j\,\rho_r(t,\vec r) 
	\simeq  a^5(t)\overline{\rho}_r(t)  \int_{V_{\rm e}}\!\!\d^3x\,x^i x^j.
\end{align}
Here and henceforth, $V_{\rm e}\equiv V_{\rm e}(t)$ will indistinctly denote both the comoving and the physical volume of the peak.
Furthermore, we have approximated the radiation energy density by its background value $\overline{\rho}_r(t)$.
In the principal axes frame, the components  of the inertia tensor are given by
\begin{equation}\label{inertia}
	\widetilde I^{ij} \equiv \int_{V_{\rm e}}\!\!\d^3x\,x^i x^j = \frac{4\pi}{15} a_1 a_2 a_3\cdot {\rm diag}(a_1^2,a_2^2,a_3^2).
\end{equation}
In general,
\begin{equation}
	I^{ij} = U^{i k} \widetilde I^{kl} U^{lj}
\end{equation}
where $U^{ij}$ is a rotation matrix.
The quadrupole then becomes
\begin{eqnarray}
	Q^{ij} =   \frac{4\pi}{45} a^5(t)\overline\rho_r(t)   a_1 a_2 a_3 
	U^{i k}U^{lj} 
	\cdot
	\left(\begin{array}{ccc}
	 2a_1^2-a_2^2-a_3^2 & 0 & 0 \\ 0 & -2a_2^2+a_1^2+a_3^2 & 0 \\ 0 & 0 & 2a_3^2-a_1^2-a_2^2
	\end{array}\right)^{kl}\hskip -0.3cm.
	\end{eqnarray}
This expression clearly shows that the quadrupole vanishes in the limit of a spherical peak.

\bigskip

\noindent
\section{III. The amount of gravitational waves.}
\setcounter{equation}{0}
\setcounter{section}{3}
\noindent
In general, the GW sourced  by a time varying quadrupole is \cite{maggiore}
\begin{equation}
	h^\TT_{ij}(t, \vec r) = \frac{4 G}{c^4} \Lambda_{ij, kl} (\hat r) \int  \frac{\d^3 s}{|\vec r - \vec{ s} |}  T_{kl} \lp t-\frac{|\vec r - \vec s|}{c}, \vec s \rp ,
\end{equation}
where  $\vec r$ denotes the physical distance from the source ($\hat r$ being its direction), $\Lambda_{ij,kl}$ the projector selecting the transverse-traceless components, $T_{kl}$ the radiation energy momentum tensor and $G$ the Newton's gravitational constant. The quadrupole radiation can then be recast into the form
\begin{align}
 h^\TT_{ij}(t,  \hat r  L) =\frac{2 G}{c^4}   \frac{1}{L}  \Lambda_{ij, kl} (\hat r) \frac{\d^2}{ \d t^2} \lp \frac{Q_{kl}}{a(t)}\rp \equiv 
 \frac{2 G}{c^4}   \frac{1}{L}  \Lambda_{ij, kl} (\hat r)  \ddot{ \mathcal{ Q}}_{kl}(t)
\end{align}
where we have momentarily assumed that the GW radiation is observed from a physical distance $a(t) L$ from the source.
Upon integrating the projectors $\Lambda_{ij, kl} (\hat r)$ over the solid angle, the power emitted in GWs is given by the expression
\begin{equation}
	{\mathscr P}_\GW= \frac{ L^2 c^3}{20G } a^2(t) \la \dot{h}_{ij}\dot{h}_{ij}\ra .
\end{equation}

We will focus on the emission of GWs over the  period extending from the time of horizon-crossing $t_\H$ until the time of turnaround $t_\ta$, for which an analytical treatment  is possible. At later stages, the evolution is non-linear and the emission of GWs, if any,  will occur through higher multipoles. We will come to this issue in the conclusions.
Therefore, our estimates are conservative lower bounds.
The corresponding  energy density of GWs at the time of production $t_\H$ and per Hubble volume $V_\H$ is given by 
\be
\label{rhoth}
\rho_{\GW}(t_\H) = \frac{1}{V_\H}\int_{t_\H}^{t_\ta} \d t {\mathscr P}_\GW
	=  \frac{3 G H^3(t_\H)}{20 \pi c^5 } a^2(t_\H) \lp t_\ta -t_\H\rp
	\left \langle  \dddot{ \mathcal{ Q}}_{ij}(t) \dddot{ \mathcal{ Q}}_{ij}(t)  \right \rangle,
\ee
where we defined the time average of the quadrupole derivatives as
\begin{equation}
	\left \langle  \dddot{ \mathcal{ Q}}_{ij}(t) \dddot{ \mathcal{ Q}}_{ij}(t)  \right \rangle
	 = \frac{1}{t_\ta - t_\H} \int_{t_\H}^{t_\ta} \d t\,  \dddot{\cal Q}_{ij}(t)\dddot{ \cal Q}_{ij}(t).
\end{equation}
The emitted energy density today will be obtained by  rescaling Eq. \eqref{rhoth} like $a^{-4}$ as is appropriate for relativistic degrees of freedom.

The typical comoving  frequency  of the GWs  at the time of production can be deduced from the characteristic momentum $k_\star$ of the power spectrum. It is of the order of the Hubble rate when the  overdensity peaks reenter the horizon,
\be
f_\star \simeq  \frac{ k_\star}{2\pi}= \frac{  a(t_\H)H(t_\H)}{2\pi}= 6 \cdot10^{-9}  \lp \frac{M_H}{M_\odot} \rp^{-1/2}  \ \rm Hz .
\ee
Here, $M_H$ is the mass contained within the Hubble volume at horizon crossing.
Notice that, since  the  characteristic physical frequency of the gravitational waves emitted is  $\sim H$, the time interval over which we average is larger than the characteristic period of the emitted GWs. 

In order to find the time evolution of the contracting ellipsoid, we isolate all the time dependent factors in the physical positions and write 
\begin{equation}
\vec r(t)= a(t) \llp 1+ g(t)\rrp \vec  q, 
\end{equation}
where we parametrise the motion given by the peculiar velocities of the collapse with a time dependent parameter $g(t)$ (with $g(t \to t_\H)=0$). Moreover, $\vec q$ is now the Lagrangian comoving position of the ellipsoid boundary and, as such, is time-independent.

Notice that, having already isolated the source of the asphericity through the different values of the semi-axes, we may  take  the evolution of the different axes to be  the same at first order in the density perturbation.

\begin{figure}
\centering
\includegraphics[width = 0.47 \linewidth]{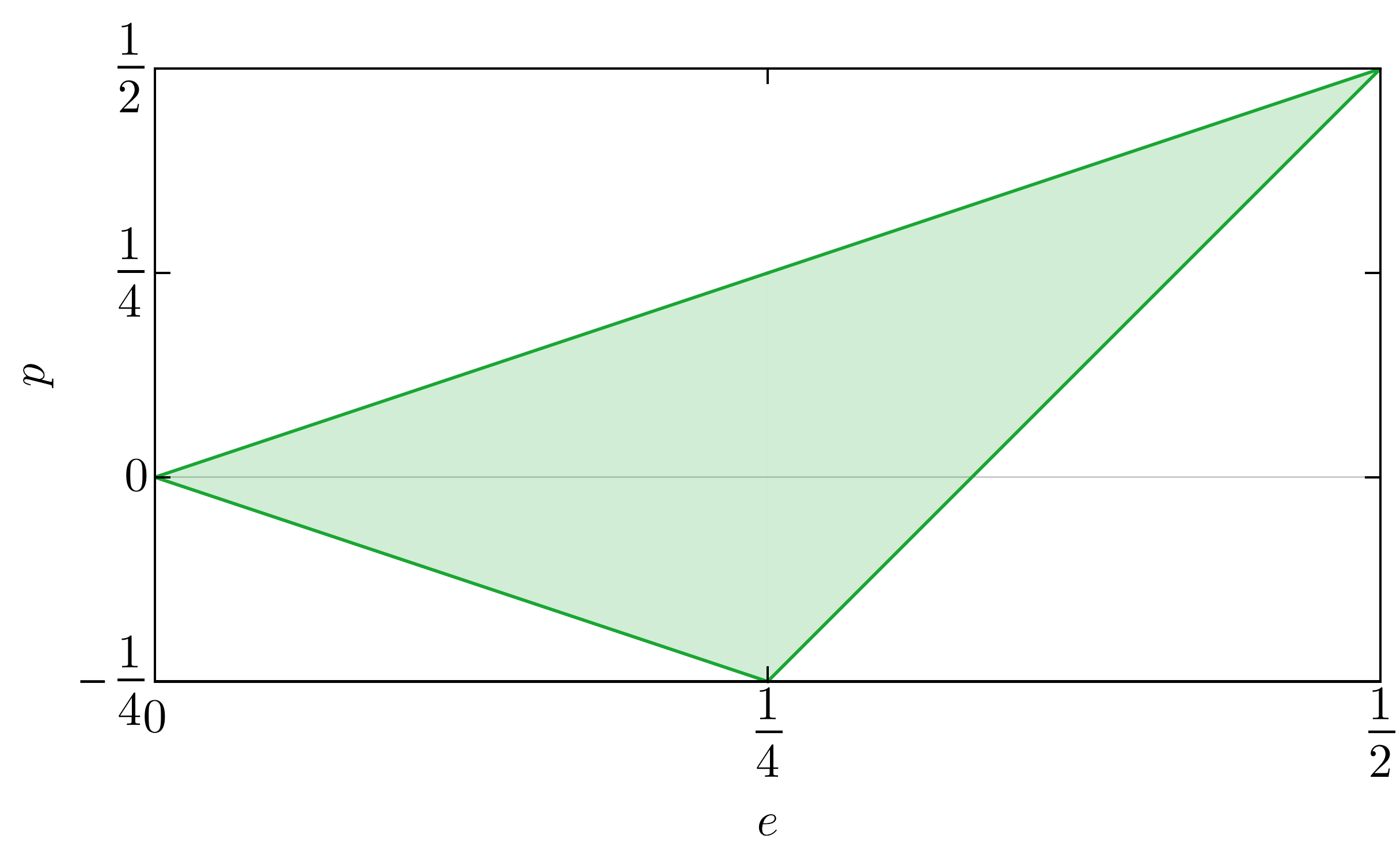}
	\includegraphics[width = 0.51 \linewidth]{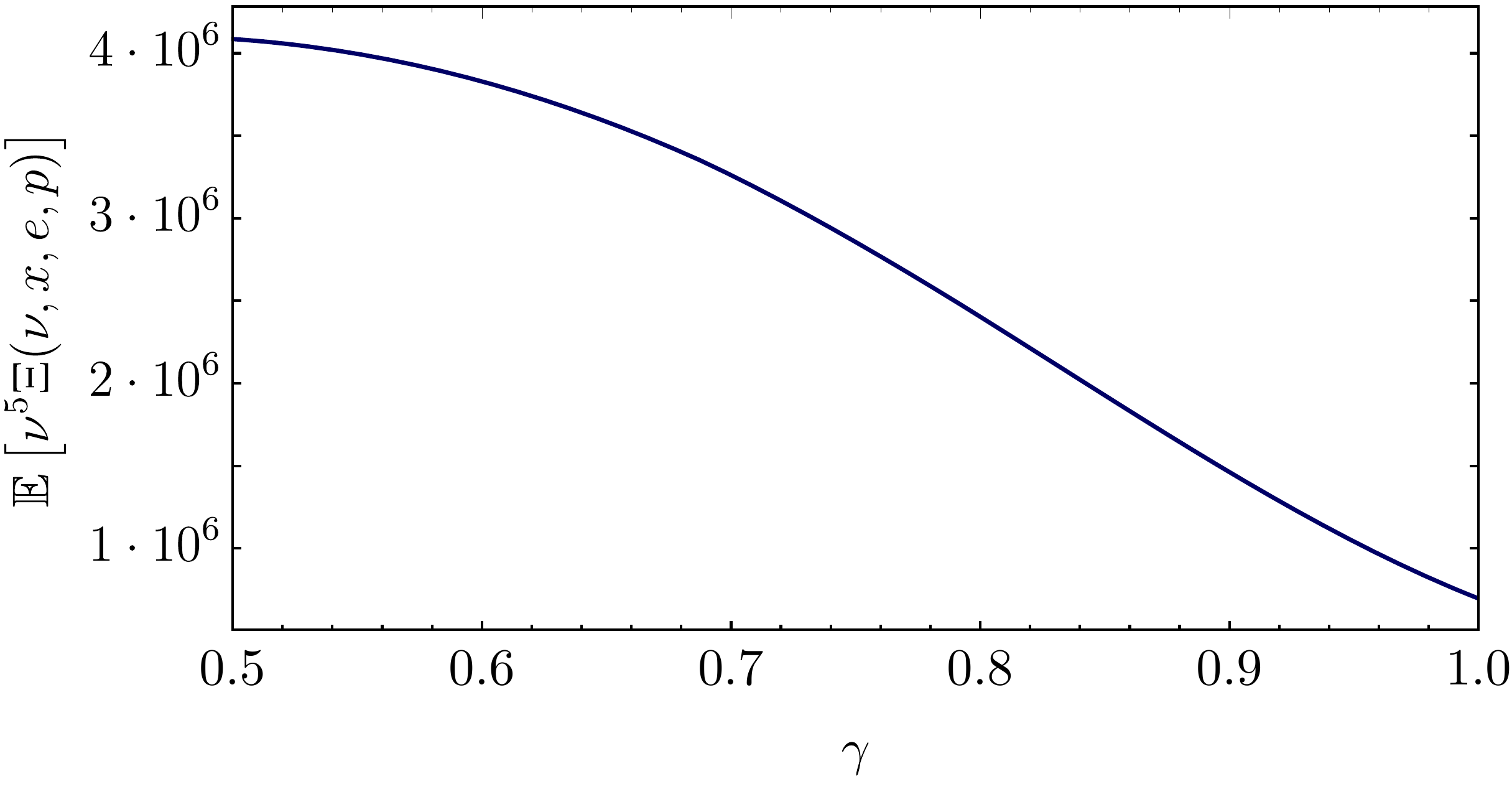}
\caption{\it Left: Support of $\chi (e,p)$ enforcing   $\lambda_1\geq \lambda_2 \geq \lambda_3 \geq 0$. Right: Expectation value $\mathbb{ E}\llp \nu^5 \, \Xi  (\nu, x , e, p) \rrp $ as a function of the parameter $\gamma$.}
\label{chiplane}
\label{rescaled quadrupole}
\end{figure}

Since the initial expansion with the Hubble flow cannot lead to any GWs emission, we must extract the quadrupole time dependence induced by the peculiar motion. For this purpose, we write
\begin{align}
	\left \langle  \dddot{ \mathcal{ Q}}_{ij}(t) \dddot{ \mathcal{ Q}}_{ij}(t)  \right \rangle
	 &=
{\rm Tr}	\left\{  \llp  
\int_{ V_{\rm e}} \d ^3  q \lp  q^i q^j -\frac{1}{3} \delta^{ij} q^2\rp \rrp^2 \right\}
	   \frac{1}{t_\ta - t_\H} \int_{t_\H}^{t_\ta} \d t\, 
\left [ \frac{\d ^3 }{\d t^3	} \left  (\overline{\rho}_r(t)  a^4(t) g^5 (t)      \rp \right ]^2
\nonumber \\
&
= 
\llp	\frac{512\pi^2}{2025}     \left(\frac{\sigma_0}{\sigma_2} \right)^5
\Xi  (\nu, x , e, p) \rrp 
 \frac{1}{t_\ta - t_\H} \int_{t_\H}^{t_\ta} \d t\, 
\left [ \frac{\d ^3 }{\d t^3	} \left  (\overline{\rho}_r(t)  a^4(t) g^5 (t)      \rp \right ]^2,
\end{align}
where  one has to keep in mind that the combination $ \overline \rho_r(t) a^4(t)$ is constant in a radiation dominated universe, so that the time derivatives only act on $g(t)$. Furthermore, we have defined a function $\Xi (\nu, x , e, p )$ encoding the geometry of the ellipsoidal peak as
\begin{equation}
\frac{\Xi  (\nu, x , e, p) }{2187 } \equiv 
\left ( \frac{\nu }{x}\right) ^5
\frac{9 e^4+3 e^2 \llp 2 (p-3) p+1\rrp +p^2 (p+1)^2}
{(1-2 p)^3 \left[(p+1)^2-9 e^2\right]^3}.
\end{equation}
Here  $e$ is the ellipticity,  $p$ the prolateness and $x=-\nabla^2\delta/\sigma_2$ (not to be confused with the comoving coordinate). Those three parameters encodes the information about the ellipsoidal shape in terms of the eigenvalues $\lambda_i$ as
\begin{align}	
	x=\lambda_1+ \lambda_2+\lambda_3,
	\qquad
	e=\frac{\lambda_1- \lambda_3}{2 \lp \lambda_1+ \lambda_2+\lambda_3 \rp},
	\qquad
	p=\frac{\lambda_1-2\lambda_2+ \lambda_3}{2 \lp \lambda_1+ \lambda_2+\lambda_3 \rp}.
\end{align}
 Notice again that the quadrupole vanishes
in the limit $e=p=0$, that is, when the peaks become spherical.
\\ \indent
Using the separate universe approach to describe the evolution of the overdensities in terms of the closed overdense metric in Eq.~\eqref{overmetric}, see Appendix A, the physical positions can be written as 
\begin{equation}
\vec r (t)= A(t) \vec q
\end{equation}
from which we infer the relation
\begin{equation}
g(t) = \frac{A(t)}{a(t)}-1 = - \delta(t_\H) \frac{  H^2(t_\H) (t-t_\H)^2}{4 H(t_\H) (t-  t_\H)+2} + {\cal O} (\delta^2).
\end{equation}
Therefore, we find
\begin{align}
\la \dddot{\cal Q}_{ij}(t)\dddot{\cal Q}_{ij}(t)\ra  
 \approx  \frac{512\pi^2}{2025}  a^{8}(t_\H)\overline\rho^2_r(t_\H)   \left(\frac{\sigma_0}{\sigma_2} \right)^5 
\Xi  (\nu, x , e, p)	\frac{1}{t_\ta - t_\H} \llp 7 \cdot 10^{-4}\cdot \delta^5(t_\H) H^5(t_\H) \rrp ,
\end{align}
where $\delta(t_\H) = \nu \sigma_0$.
In order to have a feeling of the numbers involved, one can estimate the typical value of the overdensity $\delta(t_\H)$ from the relation  
\begin{equation}
\delta (t_\H)= \nu  \sigma_0  \simeq 0.7 \cdot 
\lp \frac{\sigma_0 }{0.2} \rp,
\end{equation}
where we have taken the reference value $\nu=3.5$ and we have chosen the variance such that it yields a significant fraction of PBHs of masses $M \sim M_\odot$. 
 The time average of the third derivatives of the quadrupole is finally given by
 \begin{align}
 \la \dddot{\cal Q}_{ij}(t)\dddot{\cal Q}_{ij}(t)\ra 
=\frac{512\pi^2}{2025}  a^{8}(t_\H)\overline\rho^2_r(t_\H) H^5(t_\H)  
\left(\frac{\sigma_0}{\sigma_2} \right)^5 \frac{1}{t_\ta - t_\H} \lp 7 \cdot 10^{-4}\cdot \sigma_0^5  \rp \nu^5 \Xi  (\nu, x , e, p),
\end{align}
where the factor $\nu^5 \Xi  (\nu, x , e, p)$ depends on the geometrical  parameters $\nu$, $x$, $e$ and $p$ which describe the rescaled ellipsoidal peaks. 
Assuming gaussian statistics, the comoving number density probability distribution of these geometrical parameters reads \cite{bbks}
\begin{align}
 {\cal N}_\pk (\nu,x,e,p) 
= 
\frac{225 \sqrt{5}}{8 \pi ^3}\frac{1}{  R_*^3} \frac{1}{\sqrt{1-\gamma ^2}} x^8 e (2 p-1)\left(e^2-p^2\right) \left(9 e^2-(p+1)^2\right)
 \exp
\left[ -\frac{1}{2}\nu ^2-\frac{5}{2} (3e^2+p^2) x^2-
\frac{(x-\gamma  \nu )^2}{2(1-\gamma ^2)}\right]
\end{align}
where 
\be
R_*=\sqrt{3}\frac{\sigma_1}{\sigma_2}\,\,\,\,{\rm and} \,\,\,\,\gamma=\frac{\sigma_1^2}{\sigma_0\sigma_2}.
\ee
The parameter $\gamma$ is a measure of the width of the power spectrum: a very narrow power spectrum has a $\gamma$ close to unity.
We can integrate over this probability distribution in order to get the expectation value for the geometrical dependent factor $\nu^5 \Xi  (\nu, x , e, p)$:
\begin{align}
\label{exp}
\mathbb{ E}\llp \nu^5 \, \Xi  (\nu, x , e, p) \rrp \equiv 
V_\H
 \int_{1}^{\infty} \d \nu \int_{0}^{\infty}  \d x \int \chi(e,p)  \d e \d p\,  {\cal N}_\pk (\nu,x,e,p) \,\nu^5 \, \Xi  (\nu, x , e, p),
\end{align}
where \cite{bbks}
\begin{equation}
\chi (e,p) = \hskip -0.1cm
\left \{
\begin{aligned}
&1 \quad \text{if} \quad 0 \leq e \leq 1/4 \,\, \text {and}  -e \leq p \leq e,
\\
&1 \quad \text{if} \quad 1/4 \leq e \leq 1/2 \,\,\text {and}   -(1-3e) \leq p \leq e,
\\
&0 \quad \text{elsewhere},
\end{aligned}
\right.
\end{equation}
is a function which enforces the ordering $\lambda_1\geq \lambda_2 \geq \lambda_3 \geq 0$. The corresponding range of integration is shown in Fig. \ref{chiplane},
in which we also display the behaviour of this expectation value of $\nu^5 \Xi  (\nu, x , e, p)$ as a function of $\gamma$.
The integration range for the variable $x$ runs over positive values since we only consider local maxima of the overdensity.

To determine the lower bound of the integration range for $\nu$, note that our calculation applies to density peaks that can be distinguished from the small underlying fluctuations in the mass distribution. To be more quantitative, we follow \cite{bbks} and consider the average profile of the density perturbation at the distance $r$ from the center of the peak
\be
\overline\delta(r) \simeq \delta_{\rm pk} \frac{\xi(r)}{\sigma_0^2}
\ee
where $\xi(r)$ is the two-point correlation function.
 We can compute the root-mean-square deviation of the profile $\delta(r)$ from the average profile as \cite{Hoffman}
\be
\langle [\delta(r)-\overline\delta(r)]^2\rangle = \sigma_0^2 \llp 1-\frac{\xi^2(r)}{\sigma_0^4}\rrp < \sigma_0^2.
\ee
In order to have peaks that are statistically significant, we thus require 
\be
\nu \equiv \frac{\delta_\pk}{\sigma_0} \gsim 1.
\ee
Therefore, we will integrate $\nu$ over the range $\nu>1$. This actually provides a lower estimate because it ignores isolated peaks with significance $\nu\lesssim 1$, which should in principle also be accounted for.

Finally, to compute the present-day energy density $\rho_\GW(t_0)$, we start from the relation $\rho_\GW(t_0) = a^4(t_\H)\rho_\GW(t_\H)$, where 
\begin{align}
\rho_\GW(t_\H) = \frac{128\pi G}{3375 c^5}  a^{10}(t_\H)\overline\rho^2_r(t_\H) H^8(t_\H)  
\left(\frac{\sigma_0}{\sigma_2} \right)^5 \lp 7 \cdot 10^{-4}\cdot \sigma_0^5 \rp \mathbb{ E}\llp\nu^5 \,\Xi  (\nu, x , e, p) \rrp,
\end{align}
in which we substitute
\begin{align}
\frac{1}{a(t_\H)}&=\frac{1}{a_{\rm eq}}\frac{a_{\rm eq}}{a(t_\H)} = \frac{1}{a_{\rm eq}} \sqrt{\frac{t_{\rm eq}}{t_\H}}
=
\frac{1}{a_{\rm eq}} \frac{a (t_\H) H(t_\H)}{a_{\rm eq} H_{\rm eq}} = \frac{2 \pi f_\star}{a_{\rm eq}^2}  \frac{1}{H_0 \sqrt{ 2 a_{\rm eq}^{-4} \Omega_{r}}}
=
\frac{2 \pi f_\star }{\sqrt{2}} \frac{1}{H_0 \sqrt{  \Omega_{r}}}.
\end{align}
We have used the relations $a\sim t^{1/2}$ and $H\sim 1/t$, which are valid during the radiation phase. Moreover, $\Omega_r$ and $H_0$ are the current radiation abundance and Hubble rate respectively, while $a_{\rm eq}$ stands for the scale factor at equality.
Taking into account that, in the Standard Model of particle physics, the effective degrees of freedom of the thermal radiation change during the cosmological evolution, we introduce a factor $c_g \simeq 0.4$ \cite{noi2} in the energy density and we finally obtain  the present-day GW energy density parameter  
\begin{align}
	 \Omega_{\textrm{\tiny GW}} h^2 = 
	 4  \cdot
	  10^{-14} \lp \frac{\sigma_0}{0.2}\rp^5
	 \lp \frac{c_g}{0.4} \rp 
	 \lp k_\star^2 \frac{\sigma_0 }{ \sigma_2}\rp^5 \mathbb{ E}\llp\nu^5 \,\Xi  (\nu, x , e, p) \rrp.
\end{align} 
This is the main result of this paper.

\section{IV. Results}
\begin{figure}
	\centering
	\includegraphics[width = 0.49 \linewidth]{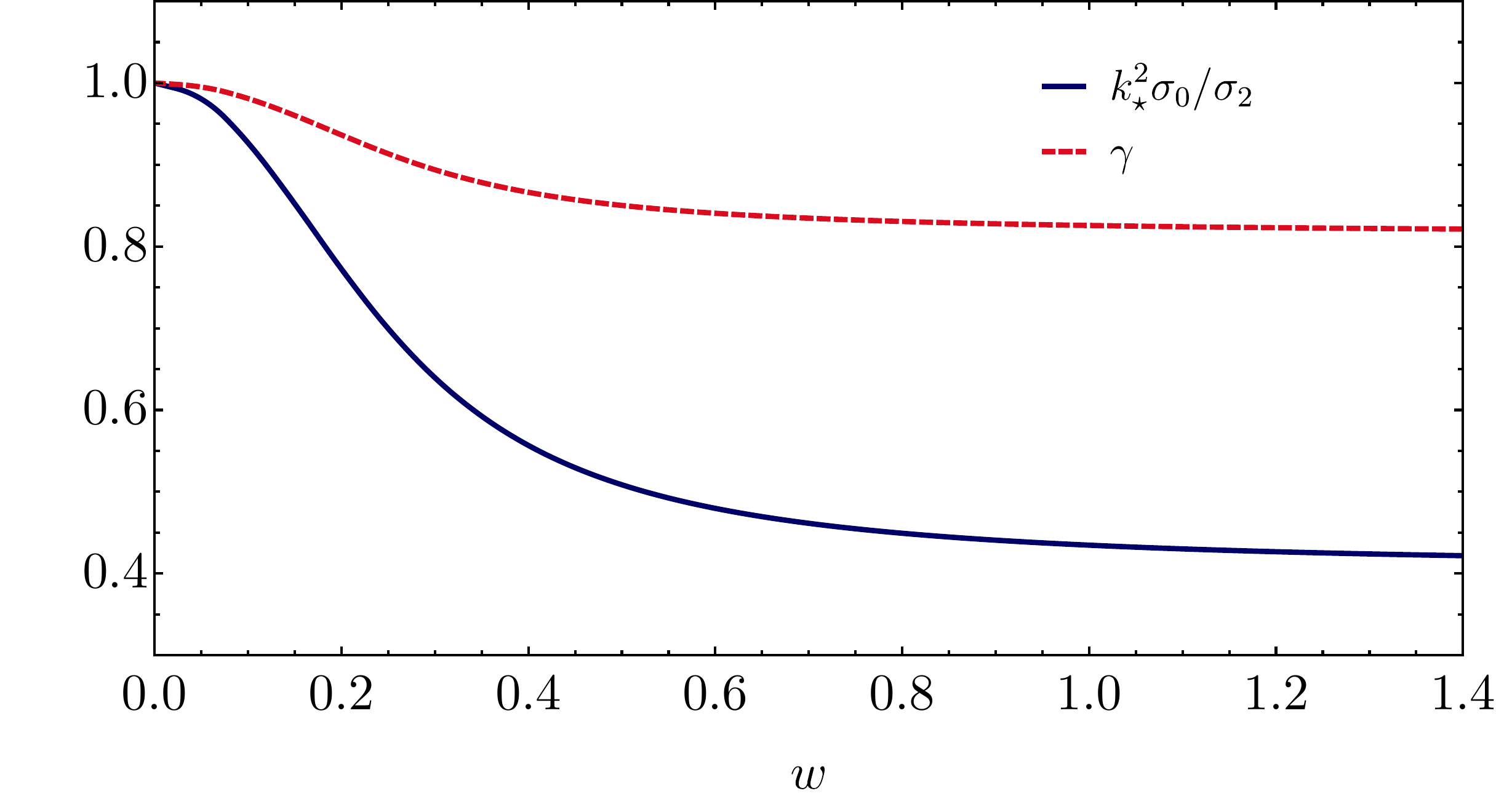}
		\includegraphics[width = 0.49 \linewidth]{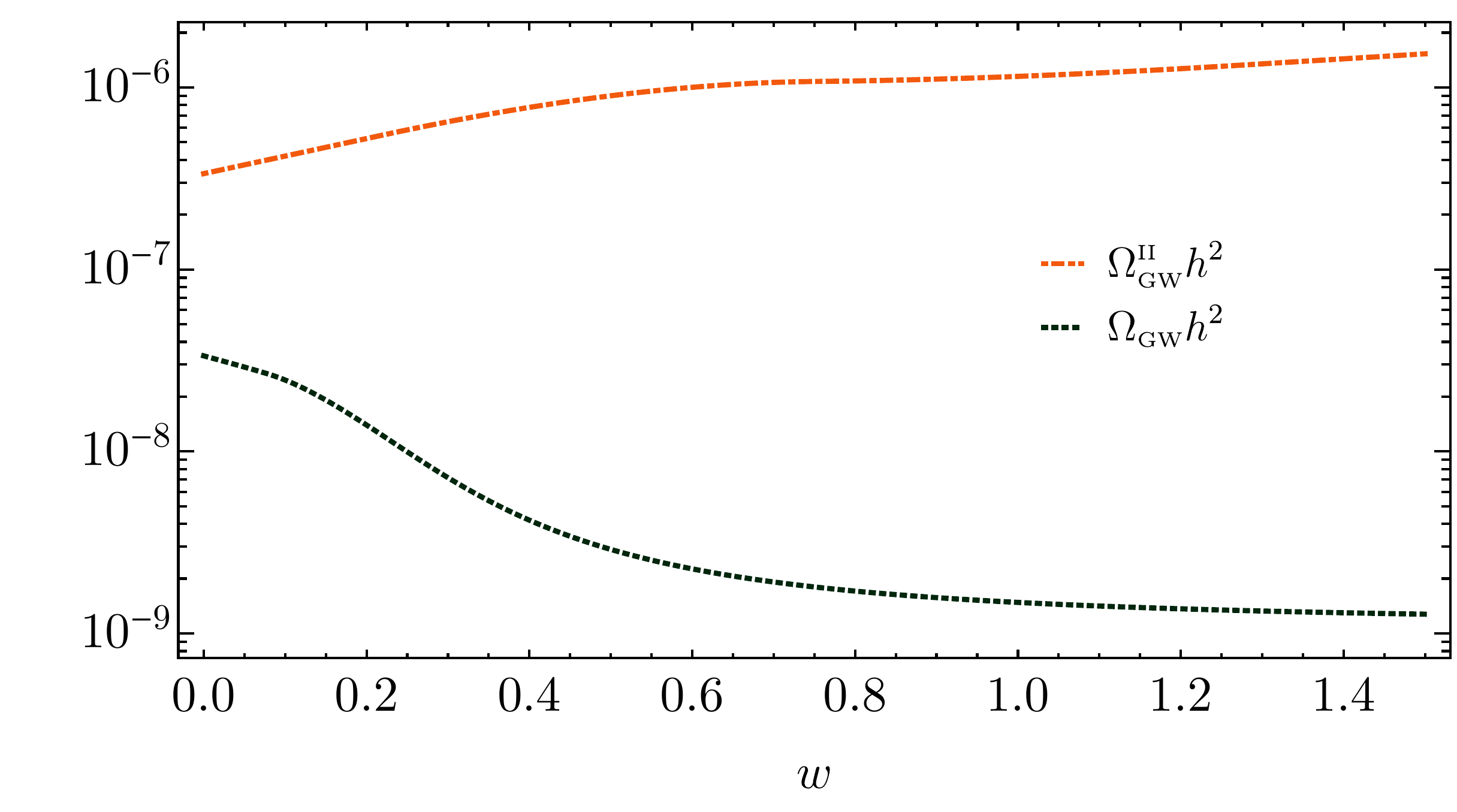}
	\caption{\it {\it Left}: The parameters $\sigma_0/\sigma_2$ and $\gamma$ dependence  on the power spectrum width $w$.
	 The two quantities reach a plateau for wide power spectra due to the presence of a window function, that here we chose to be $W(k,R_H) = \exp(-k^2R_H^2/2)$, cutting the contributions of the high-$k$ and the intrinsic cut-off at low momenta given by the gradients in the relation between $\zeta$ and $\delta$.
	{\it  Right:}  $\Omega_{\text{{\tiny {\rm GW}}}} h^2$ as a function of $w$ for a log-normal ${\cal P} _\zeta$ as in Eq.~\eqref{pslog} compared with the max abundance sourced at second order $\Omega^{\text{\tiny{\rm II}}}_{\text{{\tiny {\rm GW}}}} h^2$ by the same scalar perturbations.}
	\label{lognorm}
\end{figure}

\setcounter{equation}{0}
\setcounter{section}{4}
\noindent
To assess the importance of the GWs discussed in this paper, we assume as a representative example a   log-normal power spectrum of the comoving curvature perturbation with width $w$ and amplitude $A_g$, peaked at a characteristic scale $k_\star$
\begin{equation}\label{pslog}
	\mathcal{P} _\zeta (k) = \frac{A_g}{\sqrt{2 \pi w^2}} \exp \llp -\frac{\ln^2 (k/k_\star)}{2 w^2}\rrp.
\end{equation}
The parameters which enter our estimate of $\Omega_{\GW}$, i.e. $k_\star^2\sigma_0/\sigma_2$ and $\gamma$, depend only on the width of the power spectrum $w$ as can be seen from the left panel of Fig.~\ref{lognorm}. In the right panel of Fig.~\ref{lognorm}, we show the estimated $\Omega_{\GW}$ as a function of $w$. Notice that for $w=0$, one recovers the (unrealistic) Dirac delta power spectrum.
We also stress that the majority of the signal discussed in this paper comes from the peaks with $\nu \simeq 3.5$ due to the presence of an overall factor of $\nu^{10}$  and the Gaussian exponential in Eq. \eqref{exp}, while those
with larger values of $\nu$, which eventually end up in primordial black holes, do not contribute significantly. 

By its nature, the gravitational wave production mechanism  discussed in this paper should be regarded as complementary to the one taking place at second-order 
in perturbation theory \cite{Acquaviva:2002ud,Mollerach:2003nq,Ananda:2006af,Baumann:2007zm,Saito:2009jt,Garcia-Bellido:2017aan,errgw,noi1,noi2,Wang:2019kaf} for the same  power spectrum of the curvature perturbation with a characteristic amplitude $A_g$, whose value is fixed by the value of the variance $\sigma_0$
and set up to give enough primordial black holes to form the dark matter. In order to gauge their relative abundance, we shall compare their contributions.
We also stress the fact that the characteristic frequency of the GWs produced by the two physically different phenomena are similar, i.e. $f_\star$.
For completeness, the reader is referred to Appendix B for a concise summary of the   second-order GW source. 
The corresponding peak values of the GWs abundance  $\Omega^{\text{\tiny{II}}}_{\GW}$ are plotted in Fig.~\ref{lognorm} as a function of the power spectrum width $w$. 
The estimate  for the power spectrum  (\ref{pslog}) gives an abundance in GWs which is typically smaller than the  second-order contribution.

\begin{figure}[t]
	\centering
	\includegraphics[width = 0.7 \linewidth]{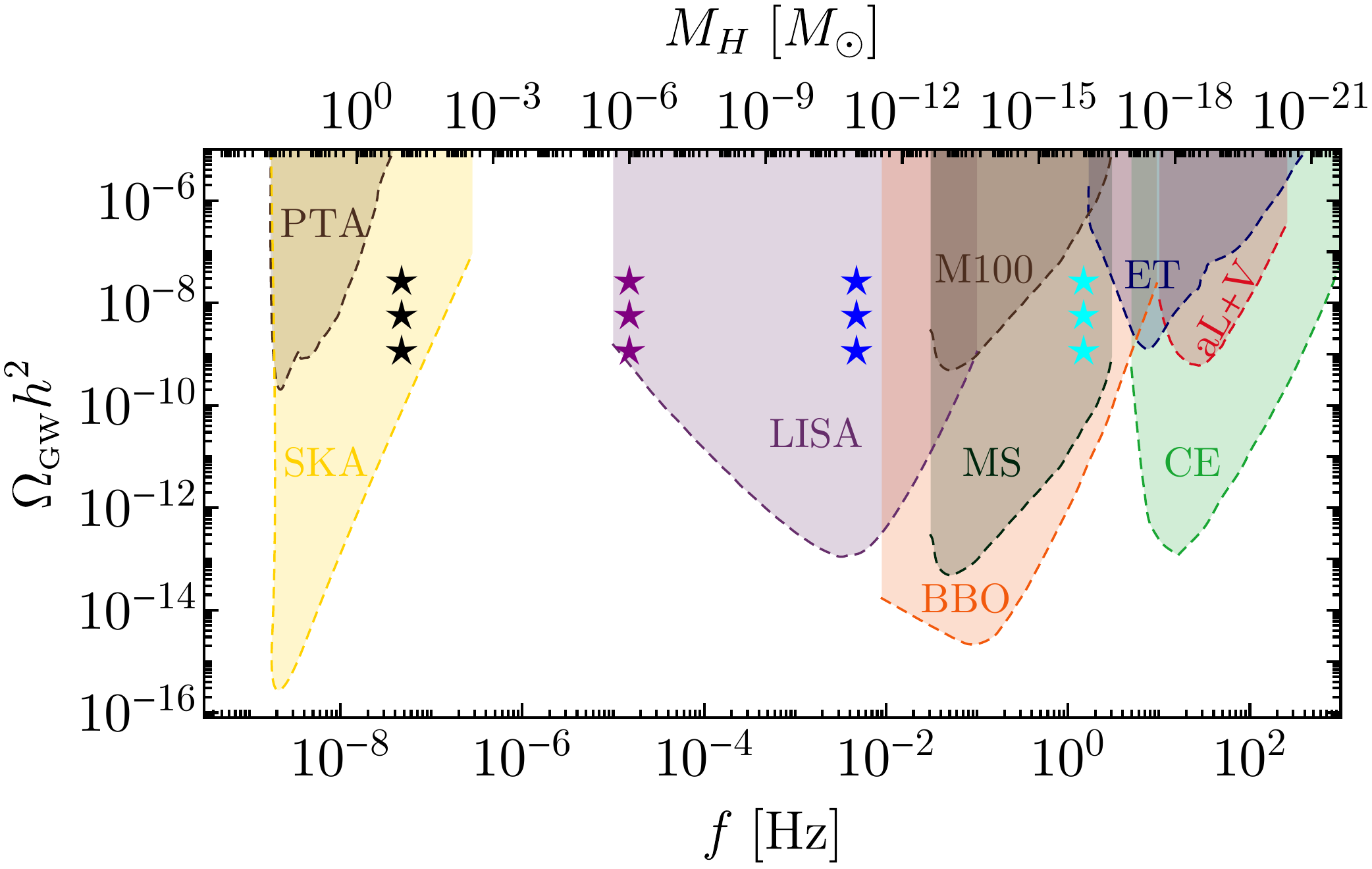}
	\caption{\it Comparison of the GWs abundance with the sensitivity curves of current/future experiments. The markers identify the  peaks of the density fraction for some representative choices of the power spectrum peak frequency $f_\star$, for the choices of $w=0$, $w=0.3$, $w=1$ respectively.
	The estimated sensitivity curves represent LISA \cite{Audley:2017drz} (power-law integrated sensitivity curve expected to fall in between the designs named C1 and C2 in Ref.~\cite{caprini}), PTA \cite{Lentati:2015qwp,Shannon:2015ect,Aggarwal:2018mgp}, SKA \cite{ska,ET-1}, DECIGO/BBO \cite{BBO}, CE \cite{Evans:2016mbw}, Einstein Telescope \cite{ET-1, ET-2}, Advanced Ligo + Virgo collaboration \cite{ligo}, Magis-AION-space \cite{magis-100} and Magis-100 \cite{magis-100}. 
}
	\label{sensitivity}
\end{figure}

In Fig. \ref{sensitivity} we have shown  the sensitivity curves of the current and future experiments and indicated by markers the corresponding     peaks of the GW abundance for representative choices of the  power spectrum peak frequency $f_\star$, for the values  $w=0$, $w=0.3$, $w=1$,  respectively. 
Despite the fact that the signal is within the reach of future experiments, it will be difficult to disentangle it from the overwhelming second order one.

We have also checked that higher-order contributions, e.g. from the octupole or the current quadrupole \cite{maggiore} are smaller than the quadrupole  contribution. The same is true for the amount of GWs coming from the angular momentum  rotation of the ellipsoidal peak induced by the action of first-order tidal gravitational fields generating first-order torques upon horizon-crossing \cite{spin}. Indeed, using the standard expressions for the GW emission from a rotating non-spherical object \cite{maggiore}, one recovers an  expression  further damped by $\omega^6\sim (v/R_*)^6$, where  $v\sim 10^{-1}\sqrt{1-\gamma^2}$ \cite{spin}.

\bigskip

\noindent
\section{V. Conclusions}
\noindent
In this paper we have discussed a possible mechanism for the generation of GWs in the early universe, which is intimately connected to the idea of producing PBHs via scalar perturbations during inflation when the latter are enhanced at small scales. PBHs are generated by the collapse of rare peaks having an amplitude larger than some critical threshold. Less prominent, albeit much more abundant peaks of lower amplitude (which do not end up in PBHs) are those playing a key role in the GW production investigated in this paper. Since they are not as spherical as the rare ones, they possess a nonvanishing quadrupole. Their density contrast grows as the Universe expands until they decouple from the Hubble flow. Subsequently, these perturbations contract, but do not form a PBH. Instead, the fluid bounces out, generating a compression wave when it encounters the surrounding plasma, giving rise to a series of bounces \cite{rez}. 

The GW production that we have studied refers to the initial stage of the overdensity dynamics, till the turnaround point, when an analytical study is possible. In the subsequent stages, the evolution is non-linear and the possible emission of GWs might occur through higher multipoles. In this sense, our estimate is conservative, but going beyond it will require a thorough numerical investigation. 
The same is true for the relation between $R_H$ and $k_\star$ which enters in the GWs abundance through the factor $(k_\star^2 \sigma_0 /\sigma_2)^5$. In particular this dependence is due to the presence of the window function in the variances, whose choice will change correspondingly the abundance of PBHs.

Our estimates indicate that the GWs from peaks would  be in principle  detectable by upcoming devoted experiments. However, their amplitude is typically smaller than the second-order  production associated
to the small-scale scalar fluctuations responsible for the birth of the PBHs and we expect this source to fully overwhelm the one discussed in the present paper.

\begin{acknowledgments}
\subsection{ Acknowledgments}
\noindent
We gratefully acknowledge I. Musco and M. Peloso for useful discussions.
A. R., V. DL. and G. F. are  supported by the Swiss National Science Foundation (SNSF), project {\sl The Non-Gaussian Universe and Cosmological Symmetries}, project number: 200020-178787.
V.D. acknowledges support by the Israel Science Foundation (grant no. 1395/16).
\end{acknowledgments}

\begin{appendices}
\section{Appendix A: Time evolution of the overdensities.}
\label{app1}
\renewcommand\theequation{A.\arabic{equation}}
\setcounter{equation}{0}
\setcounter{section}{1}
\noindent
In this appendix we report the computation of the time evolution of the overdensities until turnaround. 
On physical grounds one expects that at first stage the peaks in density are expanding following the Hubble flow. Then, after horizon crossing time $\sim t_\H$, when the internal gravitational attraction detaches them from the Hubble flow, the peaks starts contracting. Such a contraction leads to the turnaround point (at time $t_\ta$) when the velocities vanish. In this stages we are going to describe the evolution of the overdensity using the separate universe approach as in, for example, \cite{sep}.

In order to distinguish the background quantities to those belonging to the separate universe, we are going to label  $a$, $\bar \rho_r$, $H$ the former ones, while $A$, $\rho_A$, $H_A$ the latter ones. 
Locally, the metric describing a spherical region of density $\rho_A(t)=\bar \rho_r (t) \llp 1+\delta(t)\rrp $ can be written as a close FRWL metric 
\begin{equation}\label{overmetric}
	\d s^2 = - \d t ^2 + A^2 (t) \llp\frac{\d r^2}{1- K(r) r^2} + r^2 \lp \d \theta^2 + \sin ^2\theta \d \phi ^2\rp \rrp
\end{equation} 
where $K(r)$, by being a constant in time,  can be fixed at the starting point $t_\H$
\be\label{cond1}
K = \delta(t_\H)  a^2(t_\H) H^2(t_\H),
\ee
where $\delta\equiv \rho/\bar \rho_r-1$ identifies the overdensity with respect to the background.

For a closed universe, the equation of motion for the scale factor $A(t)$ is
\begin{align}\label{fried}
	H_A^2 (t) +\frac{K }{A^2(t)}  &= \frac{8 \pi G}{3} \llp \bar \rho_r (t) (1+\delta(t)) \rrp  
	= \llp 1+\delta(t_\H)\rrp H^2(t_\H) \frac{  a^4(t_\H)}{A^4(t)},
	\end{align}
since we are setting as initial conditions $A(t_\H) = a(t_\H)$, $H_A(t_\H) = H(t_\H)$ and we used the fact that in  a radiation dominated universe the energy density scales as  $\sim A^{-4}$. 

In the closed universe reference frame, the ellipsoid has fixed Lagrangian coordinates $q_i$ and physical coordinates evolving as $r_i(t) = A(t) q_i$.
Solving the Friedmann equation one gets
\begin{align}
	A(t)&=a(t_\H) \sqrt{1+ 2 H(t_\H) (t-t_\H)-\delta(t_\H) H^2(t_\H) (t-t_\H)^2 },
\end{align}
and from this solution one can find the Hubble rate to be 
\begin{equation}
	H_A(t)=\frac{H(t_\H)-\delta(t_\H)  H^2(t_\H) (t-t_\H) }{1+2  H (t_\H)(t-t_\H)- \delta (t_\H) H^2(t_\H) (t-t_\H)^2  }.
\end{equation}
The turnaround point is defined  by setting $\dot A(t_\ta)=0$ which gives
\be
t_\ta(\delta(t_\H)) = t_\H + {1}/{\delta(t_\H) H(t_\H) }.
\ee

\section{Appendix B: GWs from second-order.}
\label{app1}
\renewcommand\theequation{B.\arabic{equation}}
\setcounter{equation}{0}
\setcounter{section}{1}
\noindent
Let us remind the reader about the second-order source of the GWs.
The scalar perturbations which are responsible for the formation of peaks inevitably act also as a second-order source of GWs. 

Following the notation used in \cite{noi1}, in the Newtonian-gauge the equation of motion for the tensor perturbation can be written as 
\begin{equation}
\label{EoM-2o}
h_{ij}''+2\mathcal H h_{ij}'-\nabla^2 h_{ij}=-4 \mathcal T_{ij}{}^{\ell m}\mathcal S_{\ell m},
\end{equation}
where the primed quantities are derived with respect to the conformal time $\eta$ and $\mathcal H=a'/a = a(t) H (t)$ is the conformal Hubble parameter.
Furthermore, in Eq.~\eqref{EoM-2o} we introduced a projection tensor $\mathcal T_{ij}{}^{\ell m}$ selecting the transverse and traceless part of the source 
 given, in radiation domination (RD), by~\cite{Acquaviva:2002ud} 
\begin{equation}
\label{psi}
\mathcal S_{ij}=2\partial_i\partial_j\left(\Psi^2\right)-2\partial_i\Psi\partial_j\Psi-\partial_i\left(\frac{\Psi'}{\mathcal H}+\Psi\right)\partial_j\left(\frac{\Psi'}{\mathcal H}+\Psi\right).
\end{equation}
The scalar perturbation $\Psi(\eta,\vec k) \equiv \frac 23 T(k\eta) \zeta(\vec k) $ is related to the comoving curvature perturbation through the transfer function which, in a RD era with constant degrees of freedom, is $T(x)= (9/x^2)\left[\sin (x/\sqrt 3)/(x/\sqrt 3) -\cos(x/\sqrt 3) \right]$ ~\cite{lrreview}. The corresponding  present GW abundance is~\cite{errgw}
\begin{align}
\frac{\Omega^{\text{\tiny{II}}}_{\GW}(k)}{\Omega_{r}} &= \frac{c_g}{72}
  \int_{-1}^{1}\dd  d \int_{1}^{\infty}\hspace{-5pt}\dd  s
  \left[ \frac{( d^2-1)( s^2-1)}{ s^2- d^2} \right]^2 
{\cal I}^2\left (\frac{ d}{\sqrt{3}},\frac{ s}{\sqrt{3}} \right )  \Pz\left(k \frac{( s+  d)}{2}\right) W^2\lp k \frac{( s+  d)}{2},R_H\rp 
\nonumber \\
& \cdot    \Pz\left(k \frac{( s- d)}{2}\right) W^2\lp k \frac{( s-d)}{2},R_H \rp,
\label{eq: Omega GW with PS0}
\end{align}
where the function ${\cal I}$ accounts for a time-integrated combination of the transfer functions, see \cite{Kohri:2018awv,errgw}. 
As we assumed throughout the paper the comoving curvature perturbation $\zeta$ to be gaussian, we neglect the impact of NG corrections to the spectrum of GWs induced at second order \cite{gbp,Cai:2018dig,unal}.

\end{appendices}



\begin{thebibliography}{99}
\bibitem{ligo1} B. P. Abbott et al. [LIGO Scientific and Virgo Collaborations], Phys. Rev. Lett. 116, 061102 (2016)
  \arXiv{1602.03837}{gr-qc}

\bibitem{bird} S.~Bird, I.~Cholis, J.~B.~Mu\~{n}oz, Y.~Ali-Ha\"{i}moud, M.~Kamionkowski, E.~D.~Kovetz, A.~Raccanelli and A.~G.~Riess,
  Phys.\ Rev.\ Lett.\  {\bf 116}, no. 20, 201301 (2016)
  \arXiv{1603.00464}{astro-ph.CO}.
  
\bibitem{juan} J.~Garcia-Bellido,
  J.\ Phys.\ Conf.\ Ser.\  {\bf 840}, no. 1, 012032 (2017)
  \arXiv{1702.08275}{astro-ph.CO}.
  
 \bibitem{revPBH} 
  M.~Sasaki, T.~Suyama, T.~Tanaka and S.~Yokoyama,
  Class.\ Quant.\ Grav.\  {\bf 35}, no. 6, 063001 (2018)
  \arXiv{1801.05235}{astro-ph.CO}.
  
   
 \bibitem{s1} P.~Ivanov, P.~Naselsky and I.~Novikov,
  Phys.\ Rev.\ D {\bf 50}, 7173 (1994).
  
 
  
  \bibitem{s2} J.~Garc\'{\i}a-Bellido, A.D.~Linde and D.~Wands,
  Phys.\ Rev.\ D {\bf 54} (1996) 6040
  \arXivold{astro-ph/9605094}.

  \bibitem{s3} 
  P.~Ivanov, Phys.\ Rev.\ D {\bf 57}, 7145 (1998)
  \arXivold{astro-ph/9708224}.


\bibitem{bbks}
J.~M.~Bardeen, J.~R.~Bond, N.~Kaiser and A.~S.~Szalay,
Astrophys.\ J.\  {\bf 304}, 15 (1986).

\bibitem{ng1} M.~Kawasaki and H.~Nakatsuka,
  Phys.\ Rev.\ D {\bf 99}, no. 12, 123501 (2019)
  \arXiv{1903.02994}{astro-ph.CO}.

\bibitem{ng2} V.~De Luca, G.~Franciolini, A.~Kehagias, M.~Peloso, A.~Riotto and C.~Unal,
   JCAP {\bf 2019}, no. 07, 048 (2020)
  \arXiv{1904.00970}{astro-ph.CO}.

\bibitem{ng3}  S.~Young, I.~Musco and C.~T.~Byrnes,
  \arXiv{1904.00984}{astro-ph.CO}.

\bibitem{maggiore}
M.~Maggiore,
``Gravitational Waves: Volume 1: Theory and Experiments'',
OUP Oxford,  (2008).

 \bibitem{saito} R.~Saito and J.~Yokoyama,
  Prog.\ Theor.\ Phys.\  {\bf 123}, 867 (2010)
  Erratum: [Prog.\ Theor.\ Phys.\  {\bf 126}, 351 (2011)]
  \arXiv{0912.5317}{astro-ph.CO}.
 
 \bibitem{e}E.~Bugaev and P.~Klimai,
  Phys.\ Rev.\ D {\bf 81}, 023517 (2010)
  \arXiv{0908.0664}{astro-ph.CO}.
  
  \bibitem{gbp} J.~Garcia-Bellido, M.~Peloso and C.~Unal,
  JCAP {\bf 1709}, no. 09, 013 (2017)
  \arXiv{1707.02441}{astro-ph.CO}.


\bibitem{noi1} N.~Bartolo, V.~De Luca, G.~Franciolini, A.~Lewis, M.~Peloso and A.~Riotto,
  Phys.\ Rev.\ Lett.\  {\bf 122}, no. 21, 211301 (2019)
  \arXiv{1810.12218}{astro-ph.CO}.

\bibitem{noi2} N.~Bartolo, V.~De Luca, G.~Franciolini, M.~Peloso, D.~Racco and A.~Riotto,
  Phys.\ Rev.\ D {\bf 99} (2019) no.10,  103521
  \arXiv{1810.12224}{astro-ph.CO}.

\bibitem{spin} 
V.~De Luca, V.~Desjacques, G.~Franciolini, A.~Malhotra and A.~Riotto,
JCAP {\bf 1905}, 018 (2019)
\arXiv{1903.01179}{astro-ph.CO}.


\bibitem{Hoffman}
Y. Hoffman and J. Shaham,
Astrophys.\ J.\  {\bf 297}, (1985).



\bibitem{Acquaviva:2002ud}
  V.~Acquaviva, N.~Bartolo, S.~Matarrese and A.~Riotto,
  Nucl.\ Phys.\ B {\bf 667} (2003) 119
  \arXivold{astro-ph/0209156}.
  
  \bibitem{Mollerach:2003nq}
S.~Mollerach, D.~Harari and S.~Matarrese,
Phys.\ Rev.\ D {\bf 69} (2004) 063002
\arXivold{astro-ph/0310711}.

\bibitem{Ananda:2006af}
K.~N.~Ananda, C.~Clarkson and D.~Wands,
Phys.\ Rev.\ D {\bf 75} (2007) 123518
\arXivold{gr-qc/0612013}.

\bibitem{Baumann:2007zm}
D.~Baumann, P.~J.~Steinhardt, K.~Takahashi and K.~Ichiki,
Phys.\ Rev.\ D {\bf 76} (2007) 084019
\arXivold{hep-th/0703290}.

\bibitem{Saito:2009jt} 
  R.~Saito and J.~Yokoyama,
  Prog.\ Theor.\ Phys.\  {\bf 123}, 867 (2010)
  Erratum: [Prog.\ Theor.\ Phys.\  {\bf 126}, 351 (2011)]
  \arXiv{0912.531}{astro-ph.CO}.
 
\bibitem{Garcia-Bellido:2017aan} 
  J.~Garcia-Bellido, M.~Peloso and C.~Unal,
  JCAP {\bf 1709}, no. 09, 013 (2017)
 \arXiv{1707.02441}{astro-ph.CO}.

\bibitem{errgw} J.~R.~Espinosa, D.~Racco and A.~Riotto,
  JCAP {\bf 1809}, no. 09, 012 (2018)
  \arXiv{1804.07732}{hep-ph}.

\bibitem{Wang:2019kaf} 
  S.~Wang, T.~Terada and K.~Kohri,
  Phys.\ Rev.\ D {\bf 99}, no. 10, 103531 (2019)
  \arXiv{1903.05924}{astro-ph.CO}.

 
 
\bibitem{Audley:2017drz}
  H.~Audley {\it et al.},
  \arXiv{1702.00786}{astro-ph.IM}.

\bibitem{caprini} 
  C.~Caprini {\it et al.},
  JCAP {\bf 1604}, no. 04, 001 (2016)
  \arXiv{1512.06239}{astro-ph.CO}.


\bibitem{Lentati:2015qwp} 
  L.~Lentati {\it et al.},
  Mon.\ Not.\ Roy.\ Astron.\ Soc.\  {\bf 453}, no. 3, 2576 (2015)
  \arXiv{1504.03692}{astro-ph.CO}.
  
\bibitem{Shannon:2015ect} 
  R.~M.~Shannon {\it et al.},
  Science {\bf 349}, no. 6255, 1522 (2015)
  \arXiv{1509.07320}{astro-ph.CO}.
  
\bibitem{Aggarwal:2018mgp} 
  K.~Aggarwal {\it et al.},
  \arXiv{1812.11585}{astro-ph.GA}.
 
\bibitem{ska} 
  W.~Zhao, Y.~Zhang, X.~P.~You and Z.~H.~Zhu,
  Phys.\ Rev.\ D {\bf 87}, no. 12, 124012 (2013)
  \arXiv{1303.6718}{astro-ph.CO}.
  
 
    
  \bibitem{ET-1}
  C.~J.~Moore, R.~H.~Cole and C.~P.~L.~Berry,
  Class.\ Quant.\ Grav.\  {\bf 32} (2015) no.1,  015014
  \arXiv{1408.0740}{gr-qc}.
  
\bibitem{BBO} 
  K.~Yagi and N.~Seto,
  Phys.\ Rev.\ D {\bf 83}, 044011 (2011)
  Erratum: [Phys.\ Rev.\ D {\bf 95}, no. 10, 109901 (2017)]
  \arXiv{1101.3940}{astro-ph.CO}.
  
    
\bibitem{Evans:2016mbw} 
  B.~P.~Abbott {\it et al.} [LIGO Scientific Collaboration],
  Class.\ Quant.\ Grav.\  {\bf 34}, no. 4, 044001 (2017)
  \arXiv{1607.08697}{astro-ph.IM}.

  
  
  \bibitem{ET-2}
  B.~S.~Sathyaprakash and B.~F.~Schutz,
  Living Rev.\ Rel.\  {\bf 12} (2009) 2
  \arXiv{0903.0338}{gr-qc};
  Einstein Telescope, design at \href{http://www.et-gw.eu/}{http://www.et-gw.eu/}.
 


\bibitem{ligo}
B.~P.~Abbott {\it et al.} [LIGO Scientific and Virgo Collaborations],
 Phys.\ Rev.\ Lett.\  {\bf 118} (2017) no.12,  121101
   Erratum: [Phys.\ Rev.\ Lett.\  {\bf 119} (2017) no.2,  029901]
    \arXiv{1612.02029}{gr-qc}.
    
\bibitem{magis-100} 
  J.~Coleman [MAGIS-100 Collaboration],
  \arXiv{1812.00482}{physics.ins-det}.


\bibitem{rez} I.~Musco, J.~C.~Miller and L.~Rezzolla,
  Class.\ Quant.\ Grav.\  {\bf 22}, 1405 (2005)
  \arXivold{gr-qc/0412063}{}.

\bibitem{sep} 
  T.~Harada, C.~M.~Yoo and K.~Kohri,
  Phys.\ Rev.\ D {\bf 88}, no. 8, 084051 (2013)
  Erratum: [Phys.\ Rev.\ D {\bf 89}, no. 2, 029903 (2014)]
  \arXiv{1309.4201}{astro-ph.CO}.


  
  \bibitem{lrreview}   D.H.~Lyth and A.~Riotto,
  Phys.\ Rept.\  {\bf 314} (1999) 1
  \arXivold{hep-ph/9807278}.


 
   \bibitem{Kohri:2018awv}
K.~Kohri and T.~Terada,
 Phys.\ Rev.\ D {\bf 97} (2018) no.12,  123532
 \arXiv{1804.08577}{gr-qc}.

\bibitem{Cai:2018dig} 
  R.~g.~Cai, S.~Pi and M.~Sasaki,
  Phys.\ Rev.\ Lett.\  {\bf 122}, no. 20, 201101 (2019)
  \arXiv{1810.11000}{astro-ph.CO}.


\bibitem{unal} 
C.~Unal,
Phys.\ Rev.\ D {\bf 99}, no. 4, 041301 (2019)
\arXiv{1811.09151}{astro-ph.CO}

\end{thebibliography}
\end{document}